\documentclass[runningheads]{llncs}

\usepackage[lined,algo2e,ruled,vlined]{algorithm2e}
\usepackage{algorithmic}  
\usepackage{multirow}

\usepackage{graphicx}
\begin{document}

\title{Label-efficient Contrastive Learning-based model for nuclei detection and classification in 3D Cardiovascular Immunofluorescent Images} 
%
%
\author{Nazanin Moradinasab\inst{1}\thanks{Corresponding author: nm4wu@virginia.edu}\orcidID{0000-0003-3881-8599}
\and
Rebecca A. Deaton\inst{2}\orcidID{0000-0001-7569-1924}
\and
Laura S. Shankman\inst{2}\orcidID{0000-0003-4983-9467} \and
Gary K. Owens\inst{3}\orcidID{0000-0002-7119-9657}\and
Donald E. Brown\inst{4}\orcidID{0000-0002-9140-2632}}
%
\authorrunning{Moradinasab et al.}
%
\institute{Department of Engineering Systems and Environment, University of Virginia, Charlottesville, VA, USA \and
Laboratory of Dr. Gary Owens Cardiovascular Research Center, University of Virginia, Charlottesville, VA, USA \and Department of Molecular Physiology and Biological Physics, University of Virginia, Charlottesville, VA, USA \and School of Data Science, University of Virginia, Charlottesville, VA, USA
\email{\{nm4wu,rad5x,lss4f,gko,deb\}@virginia.edu}}
\maketitle              
\begin{abstract}
Recently, deep learning-based methods achieved promising performance in nuclei detection and classification applications. However, training deep learning-based methods requires a large amount of pixel-wise annotated data, which is time-consuming and labor-intensive, especially in 3D images. An alternative approach is to adapt weak-annotation methods, such as labeling each nucleus with a point, but this method does not extend from 2D histopathology images (for which it was originally developed) to 3D immunofluorescent images. The reason is that 3D images contain multiple channels (z-axis) for nuclei and different markers separately, which makes training using point annotations difficult. To address this challenge, we propose the Label-efficient Contrastive learning-based (LECL) model to detect and classify various types of nuclei in 3D immunofluorescent images. Previous methods use Maximum Intensity Projection (MIP) to convert immunofluorescent images with multiple slices to 2D images, which can cause signals from different z-stacks to falsely appear associated with each other. To overcome this, we devised an Extended Maximum Intensity Projection (EMIP) approach that addresses issues using MIP. Furthermore, we performed a Supervised Contrastive Learning (SCL) approach for weakly supervised settings. We conducted experiments on cardiovascular datasets and found that our proposed framework is effective and efficient in detecting and classifying various types of nuclei in 3D immunofluorescent images.

\keywords{Nuclei detection \and Classification \and Point annotations.}
\end{abstract}

\section{Introduction}
Analyzing immunofluorescent images is crucial for understanding the underlying causes of myocardial infarctions (MI) and strokes \cite{biccard2018perioperative,rickard2016associations}, which are leading global causes of death \cite{virmani2000lessons}. Clinical studies suggest that plaque composition, rather than lesion size, is more indicative of plaque rupture or erosion events \cite{libby2012inflammation,pasterkamp2017temporal}. However, determining plaque composition from 3D immunofluorescent images manually is time-consuming and prone to human error. To address these challenges, an automated algorithm is needed to detect and classify nuclei in 3D immunofluorescent images.
\begin{figure}[h]
  \centering 
  \includegraphics[width=4.2in]{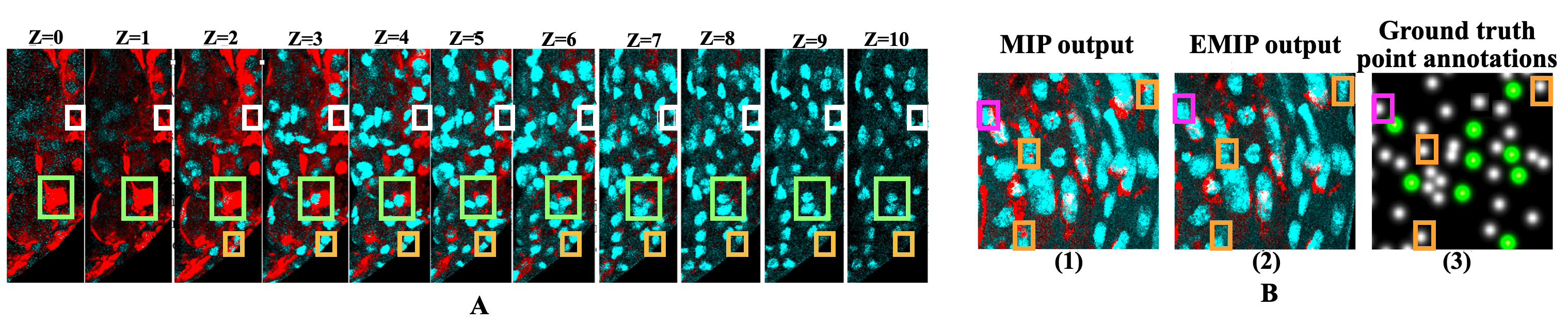} 
  \caption{A) Challenges: \textbf{(1) nucleus in the yellow square:} Even though the ground truth labels for the nucleus in the yellow and green squares are positive, they are only coincident in the second slice,  \textbf{(2) nucleus in the white square: }it shows an example of nonoverlapping marker and nucleus, (B-1) The MIP approach's output, (B-2) The EMIP approach's output, (B-3) Ground truth point annotations: green color represents nuclei labeled as positive, and white color represents nuclei labeled as negative.}
  \label{fig:challenges} 
\end{figure}
Deep learning algorithms have shown significant success in medical image segmentation and classification tasks, including histopathology and fluorescent microscopy images \cite{liu2021advances}. However, these methods typically require a large number of pixel-level annotations, which can be time-consuming and labor-intensive, especially for 3D images. To address this issue, weak-annotation methods have emerged, such as labeling each nucleus with a single point \cite{yoo2019pseudoedgenet,qu2019weakly,qu2019joint,chamanzar2020weakly,tian2020weakly,moradinasab2022weakly}. While all these studies focus on developing weakly supervised models for nuclei detection in 2D images, our study is the first to leverage weakly annotated data while preserving the 3D nature of the images.

The study conducted by Moradinasab \textit{et al.} \cite{moradinasab2022weakly} is highly relevant to our work.They introduce a weakly supervised approach for nuclei detection via point annotations and employ the HoVer-Net model. To tackle the challenge of training with point annotations, they adopt the cluster labeling approach \cite{qu2019weakly} to derive pixel-level labels.
Nevertheless, their proposed method does not classify the type of detected nuclei, which is crucial for determining plaque composition in 3D cardiovascular immunofluorescent images. Additionally, the authors converted the multichannel 3D images into z-stacked images using the Maximum Intensity Projection (MIP) technique and trained the HoVer-Net model on the resulting z-stack images. 
The MIP approach is a common technique used to reduce the computational burden in image analysis. Several studies, such as Noguchi et al. (2023) \cite{noguchi2023microscopic} and Nagao et al. (2020) \cite{nagao2020robust}, have successfully employed the MIP technique for image preprocessing to convert 3D images into 2D format in tasks like cell segmentation and cell cycle phase classification. 
Nevertheless, it's important to consider that the MIP approach might not be the optimal option for nuclei detection and classification models, as elaborated in the subsequent section.

To address these challenges, this paper proposes the Label-efficient Contrastive learning-based (LECL) model for nuclei detection/classification in 3D multichannel fluorescent images. The model aims to detect nuclei and assign them a classification label based on specific markers (e.g., Lineage Tracing) with minimum labeling cost. We assign the label "positive" if and only if the given detected nucleus overlaps with the given marker, otherwise, it is labeled as "negative". It is notable that training the model to classify the type of each nucleus using weak annotations in these images is a difficult task because these images contain multiple channels (Z-axis) for nuclei and different markers, as shown in Figure \ref{fig:nuclei_marker_linearcom}-a in Appendix. The main challenges of nuclei detection/classification in 3D images are described in detail in section 2. To address these challenges, we developed a new approach called Extended Maximum Intensity Projection (EMIP) that partially performs the maximum intensity projection per nucleus where z levels contain the given nucleus to convert multi-channel images to 2D z-stack images. Furthermore, to improve the performance of the model further in a weakly setting, we perform the supervised contrastive learning approach.

\section{Challenges}

The main challenges associated with detecting and classifying the types of nuclei in fluorescent images can be categorized into three groups as follows:

\begin{enumerate}
    \item \textbf{One specific nucleus might spread over multiple z-slices, as shown in Figure \ref{fig:challenges}-A, but only have a point annotation in one z-slice.} For example, the blue color nucleus in the orange square spreads over z-slices from two to eight, but the experts are asked to only annotate that nucleus in one of the slices to minimize the labeling cost.  
    \item \textbf{The marker and nucleus might not be coincident in all z-slices.} In fluorescent images, the given nucleus is labeled as positive if that nucleus and the marker overlap at least in one of the z-slices. In other words, even though the ground truth label for the nucleus is positive, the nucleus might not contain the marker in some slices as shown in Figure \ref{fig:challenges}-A.

    \item \textbf{Maximum Intensity Projection (MIP) can cause objects to appear coincident that are actually separated in space.} Some studies use MIP to convert multi-channel 3D images into 2D z-stack images \cite{noguchi2023microscopic}, as shown in Figure \ref{fig:nuclei_marker_linearcom}-b in Appendix. This approach utilizes MIP over nuclei/marker channels to convert these 3D images to 2D images (i.e., collapse images along with the z-axis). Then, the 2D nuclei image is combined with the 2D marker image using the linear combination method. However, this approach can be problematic when there are non-overlapping nuclei and markers in the same x and y, but at different z-axis. Figure \ref{fig:challenges}-A illustrates this, where the blue nucleus and red marker in the white square indicate non-overlapping objects that could be falsely shown as overlapping using the MIP approach.  
\end{enumerate}

 \begin{figure}[t]
  \centering 
  \includegraphics[width=4.3in]{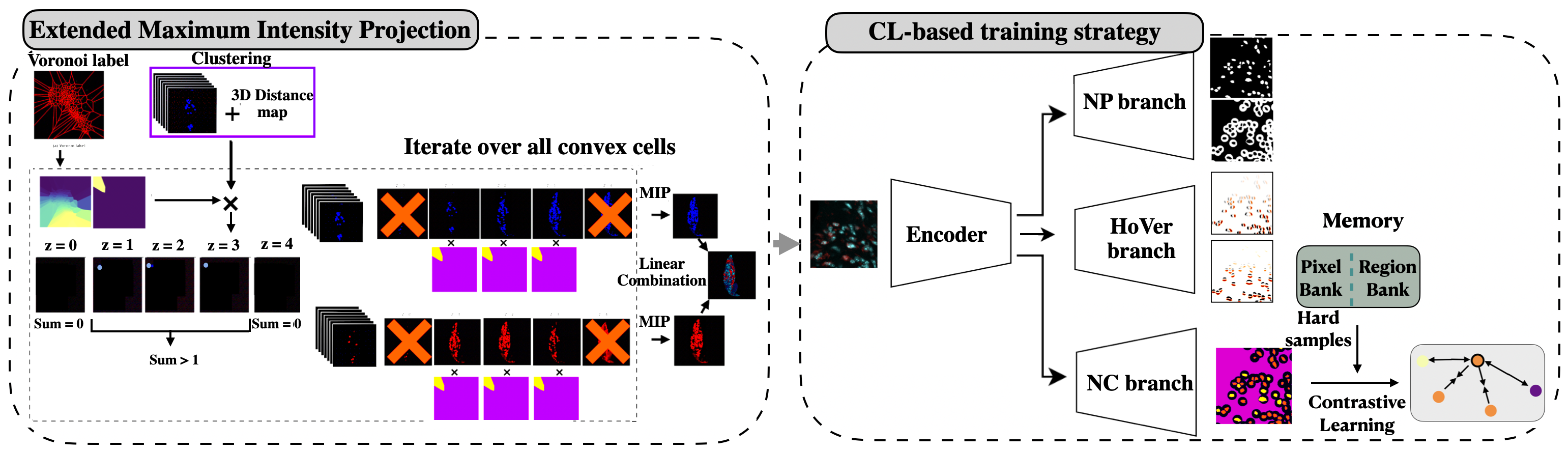} 
  \caption{Schematic representation of the LECL model}
  \label{fig:proposed_method} 
\end{figure}
\section{Method}
In this section, we describe the proposed Label-efficient Contrastive learning-based (LECL) model (Figure \ref{fig:proposed_method}), which consists of two components: a) Extended Maximum Intensity Projection (EMIP) and b) Supervised Contrastive Learning-based (SCL) training strategy.

\subsection{Extended Maximum Intensity Projection (EMIP)}
\begin{figure}[h!]
  \centering 
  \includegraphics[width=3.8in]{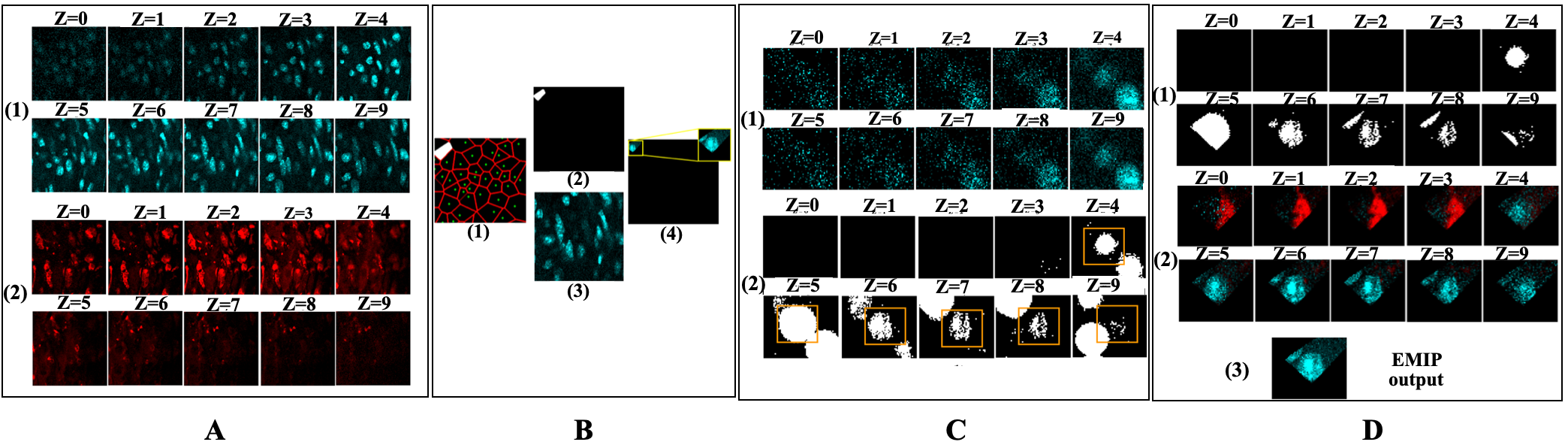}
  \caption{(A-1) The nuclei z-slices, (A-2) The marker z-slices, (B-1) The Voronoi label, (B-2) The Voronoi Cell (VC) binary mask associated to convex cell j that assigns label 1 to convex cell j and zero to others, (B-3) The z-slice 6 of nuclei channel, (B-4) The multiplication's output of VC mask and z-slice 6 which depicts the nucleus located in convex cell j, (C-1) nuclei z-slices, (C-2) The 3D binary mask, (D-1) The intersection between VC mask (B-2) and 3D binary mask (C-2), (D-2) the intersection between the VC binary mask and the nuclei/marker z-slices, (D-3) EMIP output}
  \label{fig:Sequence_nuclei_marker_example} 
\end{figure}

The EMIP method is proposed to address the issue of non-overlapping nuclei and markers in the MIP approach. It uses maximum intensity projection for each nucleus separately, only considering the z-slices containing that nucleus. For example, if a nucleus spans z-slices from seven to ten, EMIP applies MIP only to those slices. This ensures an accurate representation of the nucleus without mistakenly overlapping with the marker. Figure \ref{fig:challenges}-B compares the outputs of the MIP and EMIP approaches for an example image (Figure \ref{fig:Sequence_nuclei_marker_example}-A). The EMIP approach prevents the lineage tracing marker (in red) from falsely appearing over nuclei with the ground truth label negative. The nuclei in the pink and orange squares with the ground truth label negative falsely contain signals of the marker in the output of MIP, whereas EMIP avoids this issue.

\begin{algorithm2e}[h!]
\SetAlgoLined{linenosize=\small}
\SetKwInOut{Input}{input}
\Input{multi-channel images, Point-level annotations}
\SetKwInOut{Parameter}{parameter}
 \For{$i = 1: N$ (Number of images)}{
 \begin{enumerate}
     \vspace{-0.05cm}\item Generate the 3D distance map $(D_{i})$ using point annotations\\
     \vspace{-0.05cm}\item Create the feature map by combining the distance map and nuclei z-slices
     \vspace{-0.05cm}\item Apply k-mean clustering on the feature map 
     \vspace{-0.05cm}\item Identify background cluster (i.e., Min overlap with the dilated point labels) \\
     \vspace{-0.05cm}\item Generate 3D binary masks 
     \vspace{-0.05cm}\item Generate the 2D Voronoi label using point annotations\\
     \For{$j = 1: N_{cell}$(Number of convex cells in the 2D Voronoi label)}{
     \begin{enumerate}
     \vspace{-0.07cm}\item Generate the Voronoi Cell (VC) binary mask for cell j 
     \vspace{-0.07cm}\item Find the intersection between VC mask and 3D binary mask ($I_{j}^{3D}$) 
     \vspace{-0.07cm}\item Determine the set of slices ($S_{j}$) containing the $Nuclous_{j}$ by taking summation over the z-slices in $I_{j}^{3D}$\\
     \vspace{-0.07cm}\item Find the intersection between VC and the nuclei/marker z-slice 
     \vspace{-0.07cm}\item Compute the maximum intensity projection for nuclei and marker \\channels only over the corresponding slices ($S_{j}$) for convex cell j
     \vspace{-0.4cm}
     \end{enumerate}
     }
 \vspace{-0.9cm}
 \end{enumerate} 
     }
\caption{Extended Maximum Intensity Projection}
\label{pseudocode}
\end{algorithm2e}

Algorithm 1 outlines the steps of the EMIP approach. To perform the EMIP approach, two types of information are necessary: (a) which z-slices are associated with each individual nucleus and (b) the boundaries of each nucleus along the x- and y-axes. To determine approximate boundaries using point annotations, we propose using k-means clustering and Voronoi label methods (Steps 1-6 in Algorithm 1). It's essential to note that all nuclei have been annotated with points located at their centers, ensuring we have ground truth point annotations for all nuclei in the dataset. However, due to the nature of 3D images, nuclei are often spread across multiple slices, and the center of each nucleus is only located in one of the slices. Therefore, while every nucleus has a point annotation, these annotations are limited to the z-slice where the nucleus center is present. Consequently, point annotations for all nuclei per slice are unavailable. Using k-means clustering, we generate 3D binary masks (steps 3-5). First, we create a 3D distance map from the distance transform of point annotations (step 1). This map represents distances to the nearest nuclear point. Combining the distance map with nuclei channels of the multi-channel nuclei/marker image creates a features map (step 2). Next, k-means clustering (k=3) is applied to the feature maps, resulting in 3D binary masks. Label 0 represents the background cluster with minimal overlap with dilated point labels, and label 1 corresponds to nuclei pixels. An example of the nuclei channels and the 3D binary mask is shown in Figure \ref{fig:Sequence_nuclei_marker_example}-C. 
As shown, the binary mask indicates that the given nucleus in the orange square is spreading only over z-slices from four to nine (i.e., it approximates the nucleus' boundaries over the z-axis). To find the nuclei boundaries on the x- and y-axes, Voronoi labels are created (step 6) using point annotations (Figure \ref{fig:Sequence_nuclei_marker_example}-B-(1)). Assuming that each Voronoi convex cell contains only one nucleus, the Voronoi edges separate all nuclei from each other well.
Next, we iterate through Voronoi convex cells (steps a-e) and create a Voronoi Cell (VC) binary mask for each cell, approximating nuclei boundaries on the x- and y-axes. Figure \ref{fig:Sequence_nuclei_marker_example}-B-(2) shows the VC binary mask for convex cell j. 
Since each cell is assumed to contain only one nucleus, the intersection of the VC binary mask with nuclei/marker channels reveals the nucleus in that cell (Figure \ref{fig:Sequence_nuclei_marker_example}-B-(4)). 
Likewise, the intersection of the VC and 3D binary masks will reveal only the nucleus mask (represented by the color white) within the corresponding cell (Figure \ref{fig:Sequence_nuclei_marker_example}-D-(1)). As nuclei and background take values of one (i.e., white color) and zero (i.e., black color) respectively, simply, summation over the z-slices can be used as a detector of the presence of the nucleus. If the sum is greater than one, it implies the nucleus is present in that z-slice.  
Figure \ref{fig:Sequence_nuclei_marker_example}-D-(1) shows that the nucleus corresponding to the convex cell j is spreading over z-slices four to nine, and its boundaries over the x- and y- axes can be determined using the given convex cell edges. Figure \ref{fig:Sequence_nuclei_marker_example}-D-(3) illustrates that a MIP is performed over the z-slices spanning from four to nine for the nucleus situated within the convex cell. This technique avoids the lineage tracing marker, which spreads over slices zero to three, from overlapping with the nucleus. 

\subsection{Supervised Contrastive Learning-based (SCL) training strategy}
The Hover-Net model \cite{graham2019hover} is used for nuclei detection and classification due to its strong generalizability and instance detection performance. The generated pixel-wise masks contain three regions: nuclei, background, and an unlabeled area. For training the NC branch, Equation \ref{eq:entropy} is used, employing the cross-entropy ($L^{CE}$), Dice ($L^{Dice}$), SCL ($L^{SCL}$), and entropy-minimization ($L^{entropy}$) loss functions. The Hover and NP branches were trained using the approach from \cite{moradinasab2022weakly}. The entropy-minimization loss is applied to the unlabeled area.

\begin{equation}\label{eq:entropy}
\begin{array}{l}
  L = L^{CE}+L^{Dice}+L^{SupCon}+L^{entropy}\\
  L^{CE} = -\frac{1}{n}\Sigma_{i=1}^{N}\Sigma_{k=1}^{K} X_{i,k}(I)log Y_{i,k}(I)\\[1pt]
  L^{Dice} = 1-\frac{2\Sigma_{i=1}^{N}(X_{i}(I) \times Y_{i}(I))+ \epsilon}{\Sigma_{i=1}^{N}X_{i}(I)+\Sigma_{i=1}^{N}Y_{i}(I)+\epsilon} \\
  L^{SCL}=\frac{-1}{|P(q)|}\sum_{q^{+}\in P(q)}log\frac{exp(q\cdot q^{+}/\tau)}{exp(q\cdot q^{+}/\tau)+\sum_{q^{-}\in N(i)}{exp(q\cdot q^{-}/\tau)}}\\
  L^{entropy} = -\Sigma_{i=1}^{N} \Sigma_{k=1}^{K} Y_{i,k}(I)log Y_{i,k}(I)

\end{array}
\end{equation}

where Y, X, K, N, and $\epsilon (1.0e-3)$ are the prediction, ground truth, number of classes, number of images, and smoothness constant, respectively. The Cross-Entropy Loss function has two limitations: 1) It penalizes pixel-wise predictions independently without considering their relationships, and 2) It does not directly supervise the learned representations. HoVer-Net improves upon the Cross-Entropy Loss function by incorporating the Dice loss function, which considers pixel dependencies within an image. However, the Dice loss function does not account for global semantic relationships across images. To address the issue, we enhance our model's performance by incorporating Pixel-to-Pixel and Pixel-to-Region Supervised Contrastive Learning (SCL) \cite{wang2021exploring} techniques alongside cross-entropy and Dice losses in the third branch. We introduce a projection head in the NC branch, outputting the embedding $q$ per pixel, which is optimized using the last row of Equation 1. where, $p(q)$ and $N(q)$ indicate the set of positive and negative embedding samples, respectively.

\section{Experimental Results} 

\textbf{Metrics:} To evaluate the model's performance, we utilize the popular detection/classification metrics: precision ($P=\frac{TP}{TP+FP}$), recall ($R=\frac{TP}{TP+FN}$), and F1-score ($F1=\frac{2TP}{2TP+FP+FN}$). \\

\textbf{Datasets:} We experimented with three datasets: \textbf{Cardiovascular dataset 1 (D1)} and \textbf{Cardiovascular dataset 2 (D2)}, containing advanced atherosclerotic lesion images from two mouse models. D1 has 13 images, with 11 used for training and 2 for testing. These images vary in size along the z-axis (8 to 13). We extract $256 \times 256$-pixel patches with 10\% overlap. The train and test sets have 370 and 74 patches respectively. Additionally, we have a separate evaluation set called D2 (29 images) that are used for further evaluation. Our aim was to propose a label-efficient model that achieves comparable results with minimum labeling effort, so we trained our model on the smaller dataset. Please refer to Appendix (Table \ref{tab:characteristics}) for more details. Additionally, we used the \textbf{CoNSeP dataset} \footnote[1]{https://warwick.ac.uk/fac/cross\_fac/tia/data/hovernet/} \cite{graham2019hover}, which contains 24,332 nuclei from 41 whole slide images (26 for training and 14 for testing), with 7 different classes: fibroblast, dysplastic/malignant epithelial, inflammatory, healthy epithelial, muscle, other, and endothelial. 

\textbf{Test time:} 
We combine nuclei and marker channels per slice using the linear combination method (Figure \ref{fig:nuclei_marker_linearcom}-a in Appendix). The model detects and classifies nuclei in each slice individually. The final output is integrated over the slices with this rule: If a nucleus is predicted positive in at least one slice, it is labeled positive, otherwise negative.

\begin{table}[h]
  \centering 
  \caption{The performance of proposed methods on D1 and D2}
  \begin{tabular}{|c|c|c|c|c|c|c|c|}
  \hline
    \multirow{2}{*}{\textbf{Model}} & \multirow{2}{*}{\textbf{Branch}} & \multicolumn{3}{|c|}{\textbf{D1}}&\multicolumn{3}{|c|}{\textbf{D2}} \\
    \cline{3-8}
    &&\textbf{Precision} & \textbf{Recall} & \textbf{F1} &\textbf{Precision} & \textbf{Recall} & \textbf{F1}\\
    \hline
    
    \multirow{2}{*}{HoVer-Net \cite{graham2019hover} (MIP)} & NP & \textbf{0.8898} & 0.8894 & 0.8883&\textbf{0.9233} & 0.8455 & 0.8816 \\ 
    & NC & 0.6608 & 0.8511 & 0.7424& 0.7150 & 0.6663 & 0.6703 \\ 
    \hline
    \multirow{2}{*}{HoVer-Net \cite{graham2019hover} (EMIP)} & NP & 0.8551 & \textbf{0.9353} & 0.8880& 0.9064 & 0.8743 & 0.8894 \\ 
    & NC & 0.7694 & 0.7800 & 0.7718 & 0.8114 & 0.7718 & 0.7760\\ 
    \hline
    \multirow{2}{*}{Qu et al. \cite{qu2019weakly} (EMIP)} & NP & 0.7774 & 0.8489 & 0.8084& 0.6525 & 0.877 & 0.7431 \\ 
    & NC & \textbf{0.8548} & 0.6048 & 0.6881& 0.8140 & 0.5749 & 0.6577 \\ 
    \hline
    \multirow{2}{*}{LECL} & NP & 0.8764 & 0.9154 & \textbf{0.8942}& 0.9215 & \textbf{0.877} & \textbf{0.8978} \\ 
    & NC & 0.8277 & \textbf{0.7668} & \textbf{0.7890}& \textbf{0.8392} & \textbf{0.7840} & \textbf{0.7953} \\ 
    \hline
  \end{tabular}
  \label{tab:EMIP} 
\end{table}
\begin{table}[h!]
\small
  \centering 
  \caption{The effect of SCL based training approach on the CoNSep dataset}
  \begin{tabular}{|c|c|c|c|c|c|}
  \hline
    Model &\textbf{$F_{d}$} & \textbf{$F^{e}_{c}$} & \textbf{$F^{i}_{c}$} &\textbf{$F^{s}_{c}$} & \textbf{$F^{m}_{c}$}\\
    \hline
    HoVer-Net \cite{graham2019hover} w/o SCL (Weakly) & 0.735 & 0.578 & 0.542& 0.461 & 0.147  \\ 
    \hline
    HoVer-Net \cite{graham2019hover} w SCL (Weakly) & 0.738 & 0.576 & 0.551& 0.480 & 0.212  \\ 
    \hline
  \end{tabular}
  \label{tab:CoNSep} 
\end{table}


\textbf{Results:} Table \ref{tab:EMIP} shows the performance of different approaches on D1 and D2. The first row shows the results of using regular MIP during both the training and test stages, while the second row shows the model's performance trained using EMIP. The NP branch indicates the model's performance in detecting nuclei, and the NC branch denotes the model's performance in classifying the type of detected nuclei. As observed, the EMIP approach improves precision and F1 score metrics by 16.43\% and 3.96\% on D1, respectively, indicating a decrease in false positives. 
To ensure a comprehensive evaluation of our proposed method, we have included Dataset D2 in our study. The selection of D2 was based on its larger size and representativeness, making it suitable for robust performance assessment. As observed, the proposed EMIP approach achieves higher precision, recall, and F1 scores than the MIP method. The study found that the EMIP approach reduces false positives in lineage tracing markers overlapping with nuclei. 
Furthermore, we compare the performance of the HoVer-Net \cite{graham2019hover} model with Qu \textit{et al.}\cite{qu2019weakly} on both datasets D1 and D2. Hyper-parameters for Qu \textit{et al.}\cite{qu2019weakly} was borrowed from \cite{qu2019weakly}. As observed, the HoVer-Net model \cite{graham2019hover}  outperforms Qu\textit{et al.}\cite{qu2019weakly} in both nuclei detection and classification. We investigate the benefits of combining SCL-based training and EMIP in the LECL model. The SCL loss enhances the model's performance by capturing global semantic relationships between pixel samples, resulting in better intra-class compactness and inter-class separability. On both D1 and D2, the LECL model outperforms other models. For visualization examples, refer to Figure \ref{fig:Hovernet_MIP_MIP} in the Appendix. Furthermore, the hyperparameters for all experiments have been provided in Table \ref{tab:hyper} in the Appendix.

\textbf{Ablation study:} 
To investigate further the performance of the SCL-based HoVer-Net, we evaluate the model on the ConSep dataset (Table \ref{tab:CoNSep}). Here, $F_{d}$ represents the F1-score for nuclei detection, while $F^{e}_{c}$, $F^{i}_{c}$, $F^{s}_{c}$, and $F^{m}_{c}$ indicate the F1-scores for epithelial, inflammatory, spindle-shaped, and miscellaneous, respectively. The SCL-based model achieves better performance.

\section{Discussion}
Developing an automated approach for 3D nuclei detection and classification in fluorescent images requires expensive pixel-wise annotations. To overcome this, we propose a LECL model, which includes the EMIP and SCL components. The EMIP approach improves upon the limitations of the MIP approach, while the SCL learning approach enhances the model's performance by learning more discriminative features. However, a limitation of this study is that it relies on point annotations performed by domain experts during training and testing, which can be challenging and prone to human error. To address this, future work could explore generating synthetic data with reliable labels and using domain adaptation techniques to improve performance on real-world datasets. Moreover, the study focuses solely on lineage tracing markers, leaving room for exploring the proposed method's performance on other markers, like LGALS3.

\subsubsection{Acknowledgments} This work was supported by an American Heart Association Predoctoral Fellowship (23PRE1028980), NIH R01 HL155165, NIH R01 156849, and National Center for Advancing Translational
Science of the National Institutes of Health Award UL1 TR003015. Furthermore, we would like to thank Simon Graham, Quoc Dang Vu, Shan E Ahmed Raza, Ayesha Azam, Yee Wah Tsang, Jin Tae Kwak, and Nasir Rajpoot for providing the public CoNSep dataset.

\appendix
\section{Appendix}

\begin{figure}[h]
  \centering 
  \includegraphics[width=4in]{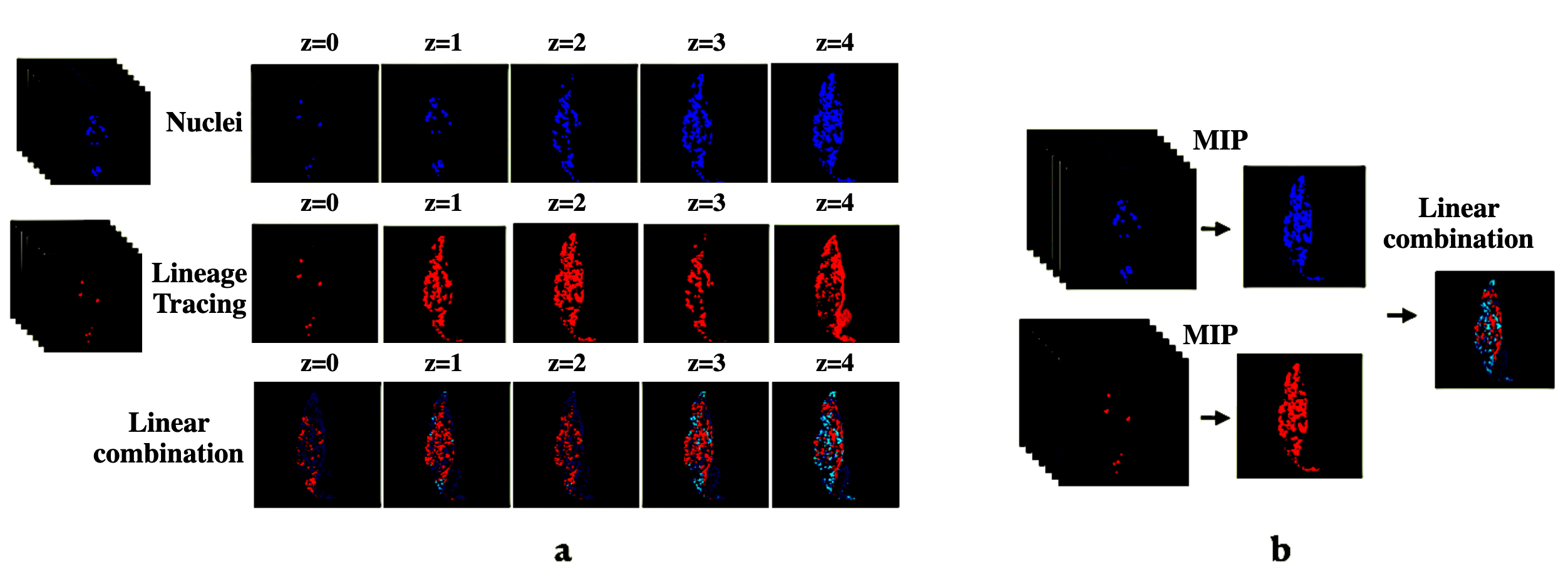} 
  \caption{a) It presents the sequence of channels (i.e. z={0,..,n}) for the nuclei (first row) and the Lineage Tracing marker (second row). The nuclei and Lineage Tracing marker channels are associated with each other in order. The third row indicates the linear combination of the nuclei and Lineage Tracing marker per slice. }
  \label{fig:nuclei_marker_linearcom} 
\end{figure}

\begin{table}[h]
  \centering 
  \caption{Cardiovascular datasets}
  \begin{tabular}{|c|c|p{0.6\linewidth}|}
  \hline
  \textbf{Charecteristices} &\textbf{Dataset}&\textbf{Value}\\ 
  \hline
    \multirow{2}{*}{\textbf{Model}} &D1&$Myh11-CreER_{T2}-RAD ROSA26-STOP_{flox}-tdTom Apoe-/- $\\
    &D2&$Myh11-CreER_{T2}-RAD ROSA26-STOP_{flox}-tdTom Irs1_{\Delta/\Delta} Irs2_{\Delta/\Delta}$ \\ 
    \hline
    \multirow{2}{*}{\textbf{Diet}} &D1&Western diet for 18 weeks\\
    &D2&Western diet for 18 weeks \\ 
    \hline

  \end{tabular}
  \label{tab:characteristics} 
\end{table}

\begin{table}[h!]
  \centering 
  \caption{Training setup for all experiments}
  \begin{tabular}{|c|c|c|c|c|}
  \hline
  \textbf{Characteristic} &\textbf{Value}&\multicolumn{2}{|c|}{\textbf{Characteristic}} &\textbf{Value}\\ 
  \hline
    \multirow{3}{*}{\textbf{Pytorch}} &\multirow{3}{*}{1.10}&\multirow{3}{*}{\textbf{Loss function weights}}&Entropy&0.5\\
    &&&Cross-entropy&1 \\
    &&&Dice loss&1\\ 
    \hline
    \textbf{GPU} &Tesla p100&
    \multicolumn{2}{|c|}{\textbf{Number of epochs}} & 100\\
    \hline
    \textbf{Projection head} &
    \multicolumn{4}{|p{0.7\linewidth}|}{\textbf{Two convolutional layers, outputting a 256 $l2$-normalized feature vector}}\\

    \hline

  \end{tabular}
  \label{tab:hyper} 
\end{table}

\begin{figure}[h]
  \centering 
  \includegraphics[width=4in]{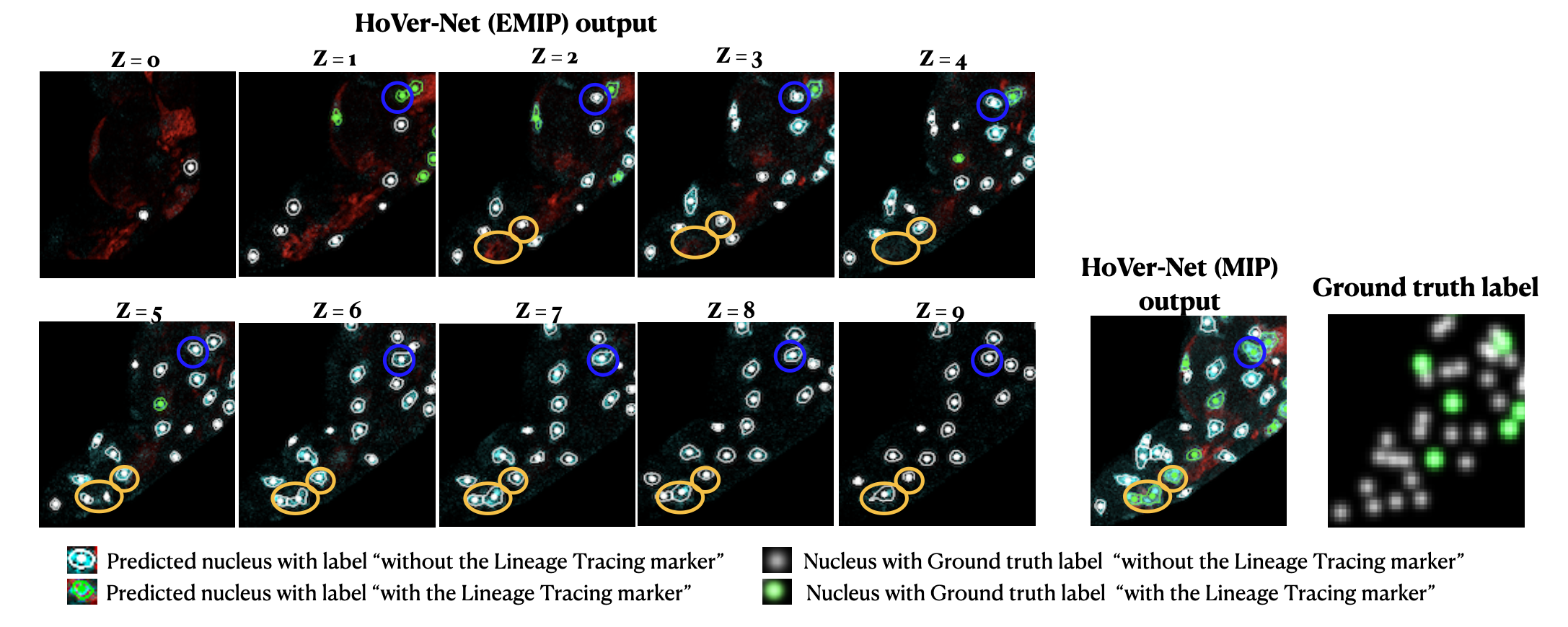} 
  \caption{The model's output trained using the HoVer-Net(MIP) and HoVer-Net(EMIP). The HoVer-Net (EMIP) model correctly predicts label positive for nuclei in the yellow circle, which is consistent with the ground truth labels. In contrast, HoVer-Net (MIP) incorrectly predicts these nuclei as negative. Both models incorrectly predict the nuclei's labels in the blue circle.}
  \label{fig:Hovernet_MIP_MIP} 
\end{figure}

\newpage

%
%
%
%
\bibliographystyle{splncs04}
\bibliography{MICCAI_workshop}

\end{document}